\documentclass[referee,useAMS,usenatbib]{mn2e}
\usepackage{graphicx}
\usepackage{subfigure}
\usepackage{natbib}
\font\twelvei = cmmi10 scaled\magstep1
       \font\teni = cmmi10 
\font\mbf = cmmib10 scaled\magstep1
       \font\mbfs = cmmib10 \font\mbfss = cmmib10 scaled 833
\font\msybf = cmbsy10 scaled\magstep1
       \font\msybfs = cmbsy10 \font\msybfss = cmbsy10 scaled 833
\def\lsim{\raise0.3ex\hbox{$<$}\kern-0.75em{\lower0.65ex\hbox{$\sim$}}}
\def\gsim{\raise0.3ex\hbox{$>$}\kern-0.75em{\lower0.65ex\hbox{$\sim$}}}

\title[Comptonized Spectra of a Neutron Star]{Monte Carlo Simulations of Thermal Comptonization Process in a Two Component Advective Flow around a Neutron Star}

\author[A. Bhattacharjee, S. K. Chakrabarti]{Ayan Bhattacharjee\thanks{ayan12@bose.res.in}$^{1}$, Sandip K. Chakrabarti\thanks{chakraba@bose.res.in}$^{1,2}$\\ 
$^{1}$ S. N. Bose National Centre for Basic Sciences, Block -JD, Sector -3, Salt Lake, Kolkata 700106, India\\
$^{2}$ Indian Center for Space Physics, 43 Chalantika, Garia St. Road, Kolkata 700084, India
}

\begin{document}

\date{}


\maketitle

\label{firstpage}

\begin{abstract}
We explore spectral properties of a Two-Component Advective Flow (TCAF) around a neutron star. We compute 
the effects of thermal Comptonization of soft photons emitted from a Keplerian disc and the boundary layer of the 
neutron star by the post-shock region of a sub-Keplerian flow, formed due to the centrifugal barrier. 
The shock location $X_s$ is also the inner edge of the Keplerian disc. We compute a series of 
realistic spectra assuming a set of electron temperatures of the post-shock region $T_{CE}$, the temperature 
of the normal boundary layer (NBOL) $T_{NS}$ of the neutron star and the shock location $X_s$. These parameters 
depend on the disc and halo accretion rates ($\dot{m_d}$ and $\dot{m_h}$, respectively) which
control the resultant spectra. We find that the spectrum becomes harder when $\dot{m}_h$ is increased.
The spectrum is controlled strongly by $T_{NS}$ due to its proximity to the Comptonizing cloud
since photons emitted from the NBOL cool down the post-shock region
very effectively. We also show the evidence of spectral hardening as the inclination angle of the disc is increased. 
\end{abstract}

\keywords{X-Rays:binaries - stars:neutron - accretion, accretion disks -
shock waves - radiation:dynamics - scattering 
}

\section{Introduction}
The theoretical modelling and spectral studies of compact objects go hand in hand in understanding the accretion processes. 
In the past twenty five years, theoretical studies of accretion onto black holes have reached a very satisfactory stage 
(for a recent review, see, Chakrabarti 2016) where it was shown that 
multiple aspects of observations of a large number of objects can be explained within a single framework. 
However, similar studies for a neutron star accretion have not been done, primarily because
it is inherently more complex. First, a neutron star can have a magnetic field which interacts with the accretion flow. If it is strong enough, it can block direct accretion and the flow then approaches 
the star through the poles. Second, it has a hard surface which can have a range of
temperatures thereby posing a difficult boundary value problem.
The matter has to stop on the surface in the co-rotating frame with the star. In other words, independent of its
past history, its final leg of journey must be sub-sonic and may form a boundary layer. On the contrary, a black hole accretion flow always 
enters through the horizon with a velocity of light and is always supersonic. Another important difference is that while for a black hole
accretion, there could be one single centrifugal barrier supported shock in a transonic flow, 
in a neutron star, there could be two shocks, one 
at a similar distance (in dimensionless unit) as in a black hole accretion and the other 
one right outside the hard surface (Chakrabarti, 1989; 
Chakrabarti \& Sahu, 1997) of the star formed due to the boundary condition. 
The outer one is known as the 
CENtrifugal barrier supported BOundary Layer (CENBOL) and the inner one may be called the Normal BOundary Layer (NBOL). 
Thus, in a black hole accretion, energy is dissipated at the CENBOL and jets/outflows 
are produced there as well. In a neutron star accretion, both the boundary layers 
can take part in dissipating the compressional heat of thermal electrons. 
So the spectral properties are more complex. 
One interesting common theme, as far as the mathematical properties go, 
is that both the flows are sub-Keplerian at the very inner edge: 
for neutron star the sub-Keplerian nature is required to adjust with the sub-Keplerian rotation of the surface; 
while for black holes, the passage through the inner sonic point ensures that the
flow is sub-Keplerian. Thus a thorough fundamental understanding of the accretion flow is important. 
The present paper is the first study to explore these aspects. 

The most complete theoretical solution in presence of advection, radiative 
transfer, and heating is a transonic or advective flow (Chakrabarti 1990; Chakrabarti 1996, hereafter, C96) 
which self-consistently passes through one or more sonic points. It was shown that in the absence of 
a significant viscosity, the flow will have a steady or oscillating shock but it would disappear
when viscosity is high and the flow would be similar to a standard Shakura-Sunyaev (1973) disc. 
Chakrabarti \& Titarchuk (1995, hereafter CT95), used the Two Component Advective Flow (TCAF) 
solution to show that spectral states of black holes could be understood by changes in the two independent 
accretion rates. Indeed, numerical simulations of Giri \& Chakrabarti (2013), 
Giri, Garain \& Chakrabarti (2015) show that the
two components are naturally produced when there is a vertical gradient of viscosity parameter. 
It was shown that when the viscosity parameter is higher than the critical value in the
equatorial region, a standard Keplerian disc is formed flanked vertically by an advective sub-Keplerian halo as 
already envisaged before (C96).

Historically, the explanation of soft state spectra of neutron stars demanded the presence of a blackbody emission from the boundary layer of a neutron star (Mitsuda et al. 1984). For the harder states with a power-law tail in the energy spectrum, the need of Compton scattering became evident (White et al. 1986, Mitsuda et al. 1989). The difference between these two models was that while the former assumed a cooler boundary layer, the latter assumed a hotter one, compared to the accretion disc. Sunyaev and his collaborators (Inogamov and Sunyaev, 1999, hereafter IS99; Popham and Sunyaev, 2001, hereafter PS01; Gilfanov and Sunyaev, 2014 hereafter, GS14) assume that the Keplerian disc reaches all the way to the neutron star and is connected with the boundary layer where the thickness increases due to higher temperature. Most of these studies were done to address the soft state spectra of neutron stars. The state transition of neutron stars in LMXBs, presented another problem. The fact that disc accretion rate was not the single factor that controlled the size or temperature of the Compton cloud, used to model the hard state spectra, lead to the conclusion that some unknown parameter, related to the truncation radius of the disc, is responsible for the hard X-ray tail (Barret 2001, Barret et al. 2002, Di Salvo and Stella 2002). Paizis et al. (2006) found a systematic positive correlation between the X-ray hard tail and the radio luminosity,  inferring that the Compton cloud might serve as the base of radio jets 
(see, Chakrabarti, 2016 and references therein). Recent phenomenological works places a transition layer or Compton cloud between the Keplerian disc and the boundary layer (Farinelli et al. 2008; Titarchuk et al. 2014, hereafter TSS14). 
It has been argued in the past (Chakrabarti, 1989; Chakrabarti, 1996; Chakrabarti \& Sahu, 1997) that
while in black hole accretion, passing of the flow through the inner sonic point ensures that the
flow becomes sub-Keplerian just outside the horizon, in the case of neutron stars, the Keplerian
flow velocity must slow down to match with the sub-Keplerian surface velocity. Numerical simulations
clearly showed that jumping from a Keplerian disc to a sub-Keplerian disc is mediated by a
super-Keplerian region (Chakrabarti \& Molteni, 1995). In Titarchuk et al. (1998, hereafter TLM98) 
a super-Keplerian transition layer was
invoked to explain the kHz Quasi-Periodic Oscillations (QPOs) and in Titarchuk et al. (2014) the transition layer 
was expanded several fold to explain the spectral properties. 
In reality, there are two such layers simultaneously present in a neutron star accretion:
One is similar to the normal boundary layer (NBOL) and the other is similar to the CENBOL in a black hole accretion (CT95).
In a black hole accretion, only CENBOL is present. All these approaches clearly point to the existence
of a CENBOL type hot electron reservoir which naturally occurred in black hole accretion, confirming Chakrabarti \& Sahu, (1997) 
conclusions that the solutions of the transonic flows are modified only in the last few Schwarzschild radii
as per the boundary condition of the gravitating object.

Monte Carlo simulations are essential in generating and understanding spectra 
emergent from highly non-local processes such as Comptonization. Toy models 
were made of spherical Compton clouds of constant temperature and optical depth, 
surrounding a weakly magnetic neutron star to generate hard X-ray tails (Seon et al. 1994). 
In case of neutron stars with strong magnetic fields ($B\sim 10^{10-12}~Gauss$), 
matter lands at the poles through the accretion column and the use of such a geometry 
leads to the successful explanation of spectral properties (Odaka et al. 2013, Odaka et al. 2014) in certain cases.
Although these studies provide some answers, so far, the spectral fitting carried out were based on phenomenological models which used arbitrarily placed Compton cloud. The TLM98 which considered a transition layer (TL)
explained high-frequency QPOs and an extended TL was used to explain spectra using COMPTT and COMPTB models. 
Out of the two COMPTB components used, the one corresponding to Comptonization of NS surface photons, 
showed a saturation in COMPTB models spectral index (Farinelli and Titarchuk 2011, hereafter FT11). 
The `Spectral slope/index' in the context of COMPTB model, refers to $\alpha$ 
which is the slope of Comptonized component of the spectrum (=$\Gamma-1$, the photon index).
In this paper, we uniformly use the actual slope of the linear region of the 
spectrum (which includes both the primary and Comptonized photons) in log-log scale. Thus the two 
terms do not represent the same quantity. Many NS LMXBs are studied using this framework, such as 
4U 1728-34 (Seifina et al. 2011), GX 3+1 (Seifina and Titarchuk 2012, hereafter ST12), GX 339+0 
(Seifina et al. 2013, hereafter STF13), 4U 1820-30 (Titarchuk et al. 2013, hereafter TSF13), 
Scorpius X-1 (TSS14), 4U 1705-44 (Seifina et al. 2015, hereafter STSS15) etc. Recently, the HMXB 
4U 1700-37 has also been examined using the same model (Seifina et al. 2016, hereafter STS16). This 
useful model to fit the spectra of neutron stars give the average temperature of the Compton cloud 
placed in between the disc and the NS surface, unlike what the TL used just outside the star. With an independent advective flow as in TCAF, the phenomenological Compton cloud is naturally explained without having to modify the earlier models drastically.
Indeed, since the flow at the outer edge of the disc has little knowledge of the nature of the compact source, except near the innermost boundary, the overall flow configuration is not expected to be very different, especially when the magnetic field is weak ($<10^8$~Gauss). The gravity simply lets the advective matter to fall almost freely till the surface of the star is hit.

In the present Paper, we study spectral properties of a neutron star in presence of TCAF which has a CENBOL as well as the normal boundary layer of the star, NBOL. A preliminary report has been presented in Bhattacharjee, Chakrabarti, \& Banerjee, (2016). We use the size of the CENBOL ($X_s$), accretion rates of the Keplerian disc 
and the sub-Keplerian halo as the free parameters. Additional parameters in the present context are the mass of the neutron 
star and the temperature of the NBOL. In the next Section, we present a brief 
introduction to the TCAF solution around a black hole and how the solution is modified when the black hole
is replaced by a neutron star. In Section 3, we describe the system in detail and define the properties of the 
NBOL, CENBOL, and the Keplerian disc. In Section 4, we describe, in brief, the Monte Carlo simulation procedure 
used in thermal Comptonization. The resultant spectra of the flow as a function of the flow parameters are 
given in Section 5. Finally, in Section 6, we give our concluding remarks.

\section{Two Component Flows Around Neutron Stars}

It is well known that the spectral and timing properties of a black hole cannot be explained by a standard Keplerian disc alone (Sunyaev \& Truemper, 1979, Sunyaev and Titarchuk 1980, 1985; Haardt and Maraschi 1993, Zdziarski 2003; Chakrabarti \& Wiita 1993; Chakrabarti \& Titarchuk, 1995; Chakrabarti, 1997). 
A spectrum clearly has a thermal component resembling of a multi-colour blackbody radiation. 
However, the other component is similar to a power-law component which is produced by inverse 
Comptonization of the thermal or non-thermal electrons (Sunyaev \& Titarchuk, 1980, 1985). There are many models in the 
literature which present possible scenarios of how the electron cloud might be produced. 
However, there is a unique self-consistent solution, namely, 
the transonic flow solution based two component advective flow (TCAF) which addresses all the aspects of 
spectral and temporal properties at the same time. In this scenario, a centrifugal barrier 
supported boundary layer or CENBOL is produced very close to a compact object in the low viscosity advection component. 
The boundary of CENBOL is a shock transition which may or may not be stationary. 
The post-shock region is a natural reservoir of hot electrons. Higher viscosity flow 
component near the equatorial plane becomes a Keplerian disc and emits a multi-colour blackbody 
radiation which are intercepted and re-radiated by the CENBOL to create the power-law like component which
normally has an exponential cut-off where recoil becomes important. When the shock oscillates due to resonance (Molteni et al. 1996) or 
non-satisfaction of Rankine-Hugoniot condition (Ryu et al. 1997), the resulting hard X-ray intensity is modulated with the size of the CENBOL and
manifest as the low-frequency quasi-Periodic oscillations (LFQPOs). The CENBOL is also the source of outflows and jets, in that, when the CENBOL is
cooled down due to excessive soft photons, the jet itself also disappears. There are clear observational evidences of two component advective flows in several black hole candidates (Smith et al. 2001; Smith et al. 2002; Debnath et al. 2013; Mondal et al. 2014; Dutta \& Chakrabarti, 2016).  
\\~
Since the nature of the companion remains generally similar irrespective of whether the compact object is a black hole or a neutron star,
it is natural that we invoke the TCAF solution for the neutron stars as well, specifically when the magnetic field is not very strong. However,
in addition to the CENBOL, to satisfy the inner boundary condition, the flow would have 
an optically thick boundary layer of the neutron star which would also emit a characteristic blackbody radiation. 
These photons would be inverse Comptonized by the electrons inside the CENBOL, cooling it down further
and making the spectra softer. In presence of a stronger magnetic field, a hot corona is produced
when matter proceeds to the magnetic poles of the neutron star. This is also the region of inverse Comptonization. 
However, discussion on this case will require inclusion of synchrotron radiation which is beyond the scope of the present paper.
In our solution, the Compton cloud is always a very hot reservoir of electrons (as compared to the NBOL or the Keplerian disc),
unless it is cooled down by the disc and/or NBOL. 

\section{Monte Carlo Simulation of TCAF around a neutron star}

\subsection{Simulation Setup}

In earlier papers from our group, Monte Carlo simulation results of two component 
advective flows around black holes were presented (Ghosh et al. 2009, 2010, hereafter, GCL09 and GGCL10, respectively). 
In the present context, in addition to the components used earlier, we must include a boundary layer (NBOL) of the star
which will emit a blackbody radiation. In Fig. 1, we present the flow configuration of our simulation. We also
use a more realistic post-shock region or CENBOL, namely, one that resembles a thick accretion disc (Molteni, Lanzafame \& Chakrabarti, 1994). 
Since we are considering a stationary configuration, to begin with, we use the density and temperature distribution as
specified by Chakrabarti (1985) inside a general relativistic model of the thick accretion disc. 

\begin{figure}
  \centering
\includegraphics[height=6.5cm,width=10.0cm]{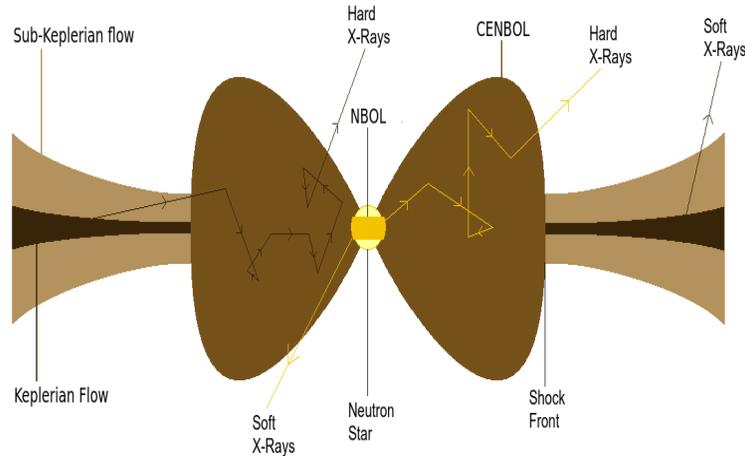}
\vskip-0.5cm
\caption{A schematic diagram of the Two-component Advective Flow (TCAF) around a neutron star. The Keplerian disc and the boundary layer on the surface of the neutron star (NBOL) emit soft blackbody photons which are scattered by the hot electrons in the centrifugal pressure supported boundary layer or CENBOL.}
\label{fig1}
\end{figure}

The three components of our simulation configuration, namely, the boundary layer of the neutron star (NBOL), the Comptonizing cloud (CENBOL) and the Keplerian disc (KD) are discussed below:

\subsubsection{\textbf{Normal Boundary Layer (NBOL)}} It is a thermal blackbody radiation emitting region on the surface of the neutron star, 
that is symmetrically placed around the equatorial plane between the azimuthal angles $(90^{\circ}-\theta_{*})~to~(90^{\circ}+\theta_{*})$, where
$\theta_{*}$ is an angle which decides the thickness of the boundary layer.
The mass of the star is kept constant at $M_{NS}=1.4~M_{\odot}$. The radius of the spherical neutron star is taken as $r_{NS}=3~r_S$, 
where $r_S=\frac{2GM_{NS}}{c^2}$ is the Schwarzschild radius of the neutron star. Here, $c$ is the speed of light in vacuum. 
From previous works carried out by IS99 and GS14 the
 temperature of an accreting neutron star is reported to be around 1.4 keV in the soft states. 
For our cases, we set the maximum value of accretion rate to one Eddington rate and the corresponding temperature 
is set to $T_{NS}^{max}=1.2~keV$. An empirical dependence on accretion rates is chosen based on the fact that 
emitted flux is proportional to the accretion rate, which is formulated as, 
\begin{equation}
T_{NS}=T_{NS}^{max}\times(\dot{m}_d+\dot{m}_h)^{1/4} {\rm keV} .
\label{tns}
\end{equation}
The flux density of photons emitted from the surface, corresponding to temperature T is calculated using (adopted from Garain et al. 2014, hereafter GGC14):
\begin{center}
\begin{equation}
n_{\gamma}=\Bigg{[}\frac{4 \pi}{c^2} \Bigg{(}\frac{k_b}{h}\Bigg{)}^3\times 1.202057\Bigg{]} T^3~cm^{-2}s^{-1},
\label{ngama}
\end{equation}
\end{center}
We follow the work of IS99, where it is reported that the angle $\theta_{*}\longrightarrow 90^{\circ}$,
when $\dot{M}\longrightarrow \dot{M}_{Edd}$. For our calculations, accretion rates are written in 
the units of $\dot{M}_{Edd}$. So, to incorporate such an effect, we assume a simpler variation of $\theta_*$, given by,
\begin{equation}
\theta_* = sin ^{-1}(\dot{m_d}+\dot{m_h}),~~~~~~ \textit{if}~ (\dot{m_d}+\dot{m_h}) \leq 1.
\label{thetastar}
\end{equation}

The flux, however, is not a constant w.r.t. $\theta$. It is believed that due to the meridional motion of the flow in the NBOL, radiation reaches a maximum flux near the angle $\theta_*$ (IS99). The effect of this is incorporated, at least qualitatively, 
by using a truncated double Gaussian distribution of photon flux where the peaks are located at $\theta_P^{\pm}$. 
We took the $95\%$ of the $\theta_*$ as the location of the peaks, as this is not crucial to the net final spectrum, i.e.,
\begin{equation}
\theta_P^{\pm} = \bigg{(}90^{\circ} \pm 0.95\times \theta_*\bigg{)}.
\label{thetapeak}
\end{equation}
The form of the distribution takes the form:
\begin{equation}
f(\theta)=exp(-(\theta - \theta_P^+)^2)+exp(-(\theta - \theta_P^-)^2).
\label{thetadist}
\end{equation}
After calculating the total flux from the entire NBOL, we redistributed the photon numbers according to the double 
Gaussian profile $f(\theta)$ along $\theta$ so as to have a realistic injection of photons from the NBOL on to CENBOL.    

\subsubsection{\textbf{CENBOL}} The centrifugal barrier supported boundary layer is modelled in shape by the equipotential contours of a standard thick disc (Chakrabarti, 1985; Chakrabarti, Jin and Arnett 1987, Ghosh et al. 2009).
The analytical forms of potential $\phi$, adiabatic constant related to entropy $K$, 
density $\rho$ and temperature $T_e$ profiles are given below:

\begin{equation}
\centering
\phi=\frac{\lambda^2}{2(r^2-z^2)}-\frac{1}{2(r-1)}=\frac{\lambda^2}{2R^2}-\frac{1}{2(\sqrt{R^2+z^2}-1)}~,
\label{phi}
\end{equation}
where $\lambda$ is the specific angular momentum, $R$ is the cylindrical radius, 
$z$ is the vertical height, $r=\sqrt{R^2+z^2}$ is the radial distance 
and $\phi$ is the specific potential energy in units of $c^2$ (GCL09; Chakrabarti 1985). 

The entropy ($K(\beta,\mu)$) is defined as (GCL09; Chakrabarti 1985),
\begin{center}
\begin{equation}
K(\beta,\mu)=\Bigg{[}\frac{3}{a}\frac{1-\beta}{\beta^4}\frac{(k_b)^4}{(\mu m_p)^4}\Bigg{]}^\frac{1}{3}~cm^3~g^{-1/3}s^{-2},
\label{entropy}
\end{equation}
\end{center}
where $\mu$ is the mean molecular weight, $a$ is the Stefan's radiation density constant, $\beta$ is the ratio of gas pressure and total pressure, $k_b$ is the Boltzmann constant, $m_p$ is the mass of the proton.

The density $\rho$ is then written as (adapted from GCL09; Chakrabarti 1985),
\begin{center}
\begin{equation}
\rho(r,z)=C_{\rho}\times\Bigg{[}\frac{\phi(r,z)}{n\gamma K}\Bigg{]}^n~g~cm^{-3},
\label{rho}
\end{equation}
\end{center}
where $n$ is the polytropic index, $\gamma$ is the adiabatic index.

The temperature $T_e$, can also be written as (adapted from GCL09; Chakrabarti 1985),
\begin{center}
\begin{equation}
T_e(r,z)=C_T\times\Bigg{[}\frac{\beta \mu m_p K}{k_b}\Bigg{]}\rho^{1/3}~K.
\label{tempe}
\end{equation}
\end{center}

For our calculations, $\lambda=1.9$, $\beta=0.5$, $n=3.0$ and the centre 
of the thick disc is at $\sim 4.25~r_S$. We did not modify the entropy by
tuning $\beta$, but kept it constant at $0.5$ throughout. Here, $C_T$ is a constant introduced to supply the central temperature as a parameter. 
In order to obtain the spectra relevant to the observed ones, we restrict the central temperature of the CENBOL to the range obtained so far 
by previous observational fits (between 3 keV to 25 keV). From the observational and theoretical studies of accretion onto 
black holes, it is well known that with increasing disc accretion rate, CENBOL would be cooled down and become smaller in size. This is regularly observed in outbursting candidates, such as GRO J1655-40, GX339-4, H1743-322, MAXI J1836-194, MAXI J1543-564 etc. (Debnath et al. 2008, 2010, 2013; Mondal et al. 2014, 2016; Jana et al. 2016; Chatterjee et al. 2016; Molla et al. 2017, 2016).
This happens because the photon flux from the disc increases with $\dot{m_d}$. This soft radiation cools down the CENBOL and 
the shock condition (balancing of pressure on both sides of the shock) is satisfied at a smaller value of shock location. 
In order to incorporate this, we use the scaling behaviour of the CENBOL with central temperature. 
For our case, a reference is set for $\dot{m_d}=0.2~\dot{M}_{EDD},~T_{CE}=10~keV,~X_s=30~r_S$. 
The density is modified by the constant $C_{\rho}$, which is determined self-consistently 
from the strong shock condition of a hybrid 1.5 dimensional advective flow solution (Chakrabarti 1989). 
The disc and halo accretion rates ($\dot{m_d}$ and $\dot{m_h}$, respectively), 
the shock location ($X_s$), and the compression ratio ($R_{comp}$) determine the 
density at the post-shock region. The pre-shock flow is assumed to 
have a velocity, $v_R \sim R^{-1/2}$ since the lower angular momentum flow would be quasi-spherical. 
In the post-shock region, the matter initially slows down and gradually picks up its speed. 
For a black hole, matter becomes supersonic before crossing the horizon. In case of neutron stars, 
the hard surface and the flow pressure force the matter to slow down just before reaching the surface. 

The parameters we use in our simulations are given in Table 1. There are altogether nine cases divided 
into three groups with different disc accretion rates (${\dot m}_d$) and the central temperatures ($T_{CE}$). Once we normalize
the outer edge of the CENBOL at $X_s=30.0$ for cases C4-C6, the outer edge changes following constant temperature 
contours and the corresponding $X_s$ are given in the Table. For each (${\dot m}_d$, $T_{CE}$) pair, we change the halo 
rate ${\dot m}_h$. Since the matter density inside the CENBOL 
is decided by the sum of the two rates, the number density of electrons at the centre of CENBOL also changes. 
The values of $T_{NS}$, $\theta_*$ and $T_e (\tau_0)$ are obtained from Eqs. 1, 3, and 13, respectively. 

In Fig. 2, we present the temperature contours corresponding to $C3$. The contours show that a CENBOL is 
essentially a toroidal star with highest density and temperature at the `centre' (which is actually a ring
around the neutron star). The vertical colour bar shows the temperatures in dimensionless unit. 
\begin{figure}
\begin{center}
\includegraphics[scale=0.4]{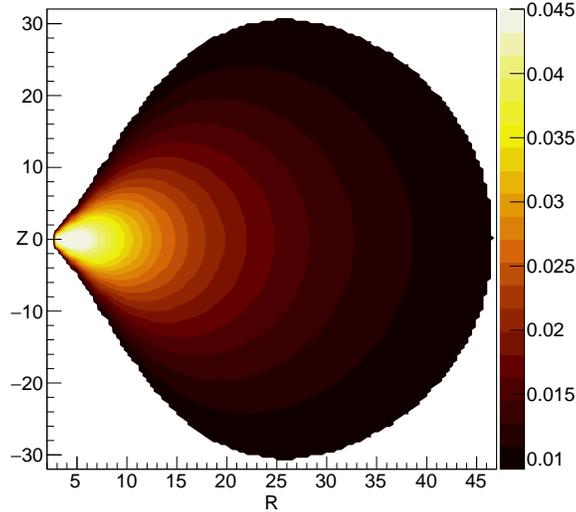}
\caption{The temperature contours used for the simulation (for $\dot{m}_d=0.1$, $T_{CE}=25.0~keV$, $X_s=46.8~r_S$. The temperatures are written in dimensionless unit $(kT_e/mc^2)$. This profile, symmetric w.r.t. z-axis, is used to get the values of temperatures at different points of CENBOL.}
\label{fig2}
\end{center}
\end{figure}

For a cylinder of half height $h_s$ and radius $X_s$, we have the accretion rate $\dot{m}_{tot}$,
\begin{equation}
\centering
\dot{m}_{tot}=(\dot{m_d}+\dot{m_h})=4\pi X_s h_s \rho_+ v_+.
\label{mtot}
\end{equation}

The density at the equatorial plane of the post-shock flow is written as (from Eq. \ref{mtot}),
\begin{equation}
\rho_+=\frac{(\dot{m_d}+\dot{m_h})R_{comp}}{4\pi h_s \sqrt{X_s}}.
\label{rhoplus}
\end{equation}
In writing Eqs. \ref{mtot} and \ref{rhoplus}, $R_{comp}$ used represents the shock compression ratio. We have assumed a strong 
shock condition for our calculation and set $R_{comp}=4$ for that purpose, throughout the simulations. As there is effectively optically 
thin matter outside CENBOL, we set the density outside to $0$ except on the equatorial plane where there is a Keplerian disc.
\begin{equation}
\rho(r,z)=0, ~~~\textit{if}~ \phi(r,z) > \phi(X_s,0).
\end{equation}
The effective temperature is calculated by transforming from coordinate space to $\tau$ space and 
then averaging over all values of $T_e(\tau)$. We followed the method of CT95 for this purpose.
\begin{equation}
T_e(\tau_0)=\frac{\int_{0}^{\tau_0}T_e(\tau)g^2(\tau)(\tau - \tau_0)^2d\tau}{\int_{0}^{\tau_0}g^2(\tau)(\tau - \tau_0)^2d\tau},~where
\label{teff}
\end{equation}
\begin{equation}
g({\tau})=\bigg{(}1-\frac{3}{2}e^{-(\tau_0 +2)}\bigg{)}\cos\frac{\pi}{2}\bigg{(}1-\frac{\tau}{\tau_0}\bigg{)}+\frac{3}{2}e^{-(\tau_0 +2)}.
\end{equation}
Here, the integration is carried out along the equatorial plane (along $R$ direction).

\subsubsection{\textbf{Keplerian Disc (KD)}} A dense region of geometrically thin and optically thick disc extends outward from 
the outer edge of CENBOL ($X_s$). The outer edge of the disk is set at $R_{out}=200~r_S$ for simplicity. The standard Shakura-Sunyaev disc 
(Shakura and Sunyaev, 1973) is used to determine the flux and height of disc as a function of radial distance in between $X_s$ and $R_{out}$. 
This acts as a secondary source of blackbody photons. Here, $dr=0.5~r_S$, $\dot{M}_{17}$ is the disc accretion 
rate in the units of $10^{17}$ gm s$^{-1}$. The radial distance $r$ is measured in the units of $r_S$. The injected energy flux is given by
(SS73; GCL09), 

\begin{center}
\begin{equation}
F(r)=5\times10^{26}(M_{NS})^{-2}\dot{M}_{17}(2r)^{-3}\Bigg{[}1-\sqrt{\frac{3}{r}}\Bigg{]}~erg~cm^{-2}~s^{-1}.
\label{fluxkd}
\end{equation}
\end{center}

From which, the effective temperature ($T(r)=(F(r)/\sigma)^{1/4}$) and height of the disc is determined as (adopted from GCL09),
\begin{center}
\begin{equation}
T(r)\approx5\times10^{7}(M_{NS})^{-1/2}\dot{M}_{17}^{1/4}(2r)^{-3/4}\Bigg{[}1-\sqrt{\frac{3}{r}}\Bigg{]}^{1/4}K,
\label{tempkd}
\end{equation}
\end{center}

\begin{center}
\begin{equation}
H(r)=10^{5}\dot{M}_{17}\Bigg{[}1-\sqrt{\frac{3}{r}}\Bigg{]}~cm.
\label{hkd}
\end{equation}
\end{center}

For our case, every cylindrical shell between $r~and~r+\delta{r}$ of height $H(r)$, emits a total $dN(r)$ number of photons per second (GGC14) from its top and bottom surfaces, where
\begin{center}
\begin{equation}
dN(r)\approx 2\pi r \delta{r} n_{\gamma}(r).
\end{equation}
\end{center}
For computational purposes, we chose packets of photon instead of individual photons, in such a way that, effective $dN(r)\sim 10^5$, which is required for a good statistics.

\section{Scattering Process}

We adopt the method followed by Pozdnyakov, Sobol, and Sunyaev (1983, hereafter PSS83) to model the thermal Compton scattering process. 

\subsection{Radiative processes}

We simulate thermal Compton scattering phenomenon between the soft photons emitted by the Keplerian disc 
(or, the NBOL, the boundary layer of the neutron star) and the hot electrons in the post-shock region (CENBOL), using a Monte Carlo code. 
Photons from the Keplerian disc and the boundary layer are modelled using Planck's distribution formula for blackbody radiation. 
The radial temperature distribution for Keplerian disc $T(r)$ is used as above.
For the neutron star, the value $(T_{NS})$ is kept constant throughout the simulation. 
In both the cases, the number density of photons ($n_{\gamma}(E)$) having an energy between $E$ and $E+dE$ is expressed by (PSS83):
 \begin{equation}
\centering
n_{\gamma}(E)=\frac{1}{2\zeta(3)}b^3E^2(e^{bE}-1)^{-1},
\label{ngamapss83}
\end{equation}
where $b=1/kT$. Here $T=T(r)$ for KD and $T=T_{NS}$ for NS. The zeta function, $\zeta(3)=\sum_{1}^{\infty}l^{-3}=1.202$.

For every packet of photon, we start the simulation by determining the local normal at the randomly chosen position on the
i) NBOL or the ii) Keplerian disc. Then, a random direction ($\theta_r$) is chosen w.r.t. the local normal, 
with the probability density $cos(\theta_r)$. We also assign a random critical optical depth $\tau_c$. 
During the photons' motion, we calculate $\tau$ by summing over $d\tau$ to check if it has crossed $\tau_c$ or not. 
Here, $d\tau =\rho_n \sigma dl$ where $\rho_n$ is the number density of electrons and $dl$ is the length traversed 
$\sigma$ is the scattering cross section, determined using the Klein-Nishina formula,

$\sigma=\frac{2 \pi r_{e}^2}{x} \bigg{[}\bigg{(}1-\frac{4}{x}-\frac{8}{x^2}\bigg{)}ln(1+x^2)+\frac{1}{2}+\frac{8}{x}-\frac{1}{2(1+x)^2}\bigg{]}$,
\\where 
$x=\frac{2E}{mc^2}\gamma \bigg{(}1-\mu \frac{\nu}{c}\bigg{)}$.
Here, $r_e = e^2/mc^2$ is the classical electron radius and $m$ is the mass of the electron.

The electrons are assumed to have a relativistic Maxwell-Boltzmann distribution of momentum. If $\textbf{p}=\gamma m \textbf{v}$, with $\gamma=(1-\frac{v^2}{c^2})^{-1/2}$ and $\mu=\hat{\textbf{$\Omega$}}.\hat{\textbf{v}}$, the number density of electrons between momentum $\textbf{p}~and~\textbf{p+dp}$ is given by PSS83, 
\begin{equation}
dN(\textbf{p}) \propto exp [-(p^2c^2 +m^2c^4)^{1/2} / kT_e]d\textbf{p}.
\label{pdist}
\end{equation}
Based on the scattering cross-section obtained, the optical depth is calculated. If, $\tau>\tau_c$, we allow the photon to scatter 
and assign a new direction based on its energy. A new random $\tau_c$ is generated and the process is continued till the photon
leaves the system under consideration. Details of the procedure is given in PSS83.

We continue the simulations for all the nine cases listed in Table 1. The angle $\theta_*$, number density $n_{ce}$ and average temperature $T_e(\tau_0)$ were all derived from other
parameters. Inclusion of the compression ratio ($R_{comp}$) and mass of the star ($M_{NS}$),
would make the number of independent parameters to be seven. For our theoretical investigations in the present paper,
we did not focus on the reduction of the number of parameters in the present work, but
we can further reduce it by interrelating CENBOL properties from fundamental equations. We vary $\dot{m}_d$ and $\dot{m}_h$ 
independently and determine the rest of the parameters either using observational facts or through the formulae 
derived in Section \S 3. This has been discussed in Section \S 5.

\begin{table}
\vskip0.2cm
{\centerline{\large Table \ref{table:parameters}}}
{\centerline{List of parameters for the system used in simulation}}
{\centerline{}}
\begin{center}
 \begin{tabular}{c c c c c c c c c} 
 \hline
 \hline
 ID & $\dot{m}_d$ ($\dot{M}_{EDD}$) & $\dot{m}_h$ ($\dot{M}_{EDD}$) & $X_s$ ($r_S$) & $T_{CE}$ (keV) & $n_{ce}~(\times 10^{18})$ & $T_{NS}$ (keV) & $\theta_*$ (degrees) & $T_{e}(\tau_0)$ (keV) \\  
 \hline
 C1 & 0.1 & 0.1 & 46.8 & 25.0 & 1.772 & 0.802 & 11.52 & 22.062 \\  
 C2 & 0.1 & 0.2 & 46.8 & 25.0 & 2.658 & 0.888 & 17.48 & 22.062\\
 C3 & 0.1 & 0.5 & 46.8 & 25.0 & 5.317 & 1.056 & 36.90 & 22.062\\
 C4 & 0.2 & 0.1 & 30.0 & 10.0 & 1.515 & 0.888 & 17.48 & 8.909\\
 C5 & 0.2 & 0.2 & 30.0 & 10.0 & 2.020 & 0.954 & 23.61 & 8.909\\
 C6 & 0.2 & 0.5 & 30.0 & 10.0 & 3.535 & 1.098 & 44.40 & 8.909\\
 C7 & 0.5 & 0.1 & 21.8 & 3.0 & 2.100 & 1.056 & 36.90 & 2.705\\
 C8 & 0.5 & 0.2 & 21.8 & 3.0 & 2.450 & 1.098 & 44.40 & 2.705\\
 C9 & 0.5 & 0.5 & 21.8 & 3.0 & 3.500 & 1.200 & 90.00 & 2.705\\
 \hline
\end{tabular}
\caption{Set of parameters chosen for the simulations. 
Disc accretion rate ($\dot{m}_d$) and halo accretion rate ($\dot{m}_h$) are varied independently. 
Temperatures are from typical observational range
and consistent with high disc accretion rates leading to lower temperatures of CENBOL. 
Shock locations ($X_s$) are chosen accordingly.
Neutron star's (NBOL) temperature ($T_{NS}$), $\theta_*$ and central number 
density $n_{ce}$ are derived using empirical rules given in the text.}
\label{table:parameters}
\end{center}
\end {table}

\subsection{Photoelectric absorption}

Photoelectric effect due to gases present in the interstellar media (ISM) is 
the most significant cause of absorption of photon, in the domain of energy we are considering in this paper. 
Depending on the degree of absorption, the observed spectra can be largely modified between the 
energies $0.03-10$ keV (Morrison and McCammon 1983). We have used the standard formula for photoelectric cross-section,
\begin{equation}
\sigma(E)=4\alpha\sqrt{2}z^58\pi\frac{r_e^2}{3}\Big{[}\frac{m_ec^2}{E}\Big{]}^{3.5}~cm^2.
\end{equation}

An exponential absorption model was used, following the `wabs' model used in XSPEC, which can be written as,
\begin{equation}
M(E)=exp(-n_H \sigma(E)),
\end{equation}
where $n_H$ is the equivalent column density of hydrogen in ISM along the observational direction.

In Figs. \ref{fig6} to \ref{fig8}, the dotted lines show our derived spectra before the interstellar absorption and 
solid lines show the spectra after absorption. For the cases C1 to C9, $n_H=4.0 \times 10^{22}~cm^{-2}$ was chosen for
concreteness.

\subsection{Effects of cooling}
We focus on the cases when the CENBOL has temperature up to 25 keV, and the accretion rates are not
 too high ($\leq \dot{M}_{Edd}$). In the flaring branches, however, the observed spectra shows the Comptonized spectrum of
to extend beyond 200 keV, pointing towards a hotter outer CENBOL. In order 
to check this effect of cooling we take the case of comparatively hotter central temperature $(T_{CE}=250~keV)$,
 and vary the accretion rates to observe the effects of cooling due to Comptonization. In the scenario 
when cooling is efficient, the static background temperature distribution should be modelled by the 
modified temperature, rather than the unmodified thick disc distribution. However, the spectra 
of those cases are not discussed any more as the Comptonized spectra of the disc is well understood in
 the cases of black hole under the TCAF paradigm. 

First, we project the entire 3D simulation region onto an axisymmetric cylindrical grid, for 
which the thick disc distributions are used. The number density of electron at each grid location 
$(ir,iz)$, which corresponds to the position $(R,z)$, is given by $n_e(ir,iz)$. Assuming a torus, 
for axisymmetric system, the total number of electrons within that volume is given by $dN_e(ir,iz)$,

\begin{equation}
dN_e(ir,iz)=2\pi R n_e(ir,iz) dR dz
\end{equation}

where $dR$ and $dz$ are the grid sizes in R and z directions, respectively.

Depending on the actual photon flux and the number of bundles of photons injected, a weightage can be assigned to each bundle. Let the weightage for $i^{th}$ packet be $f_W^i$. For a completely relativistic hydrogen plasma, the total thermal energy of all the electrons in the torus with temperature $T_e(ir,iz)$ is given by,

\begin{equation}
3k_B T_e(ir,iz)dN_e(ir,iz)=6\pi k_B R T_e(ir,iz) n_e(ir,iz) dR dz .
\end{equation}

If the $i^{th}$ packet of photon undergoes Compton scattering at the position $(ir,iz)$ on the grid
 gaining (or losing) energy $\Delta E^i$, then the final temperature $k_B T_e^{\prime}(ir,iz)$ after all 
the photons have left the system, can be written as,

\begin{equation}
k_B T_e^{\prime}(ir,iz)=k_B T_e(ir,iz)-\frac{\sum\limits_{i=1}^{i_{max}}\Delta E^i f_W^i}{3dN_e(ir,iz)}
\end{equation}

where $i_{max}\sim 10^8$.

We divide the CENBOL into two domains of equal optical depth, calculated along the equatorial plane. If the total optical depth of the CENBOL is $\tau_0$, the region up to $\tau=\tau_0/2$ from the NBOL (or the shock location $X_s$) is the cloud ``visible" to the NBOL (or the Keplerian disc), which would produce effectively the same number of scatterings. We determine the average temperature based on the method specified in eqn. 13, for both the cases, to study the effect of cooling of the CENBOL, by the NBOL and by the disc. This is just to find out the two temperatures in the two halves which Seifina et al. (STF13, STSS15) and Titarchuk et al. (TSS14) found in their work by fitting COMPTB1 and COMPTB2 respectively. The results of this excercise is shown in  Fig. 10(a-d) (below).

\section{Results}

The photons are collected outside a sphere of radius $R_{out}$ where they leave our system. They 
are binned according to their energy, angle of observation, the original location of emission (NBOL or KD) and the 
number of scatterings each of them suffered. We generate the output spectra from these informations.

\begin{center}
\begin{figure}
\includegraphics[scale=0.35]{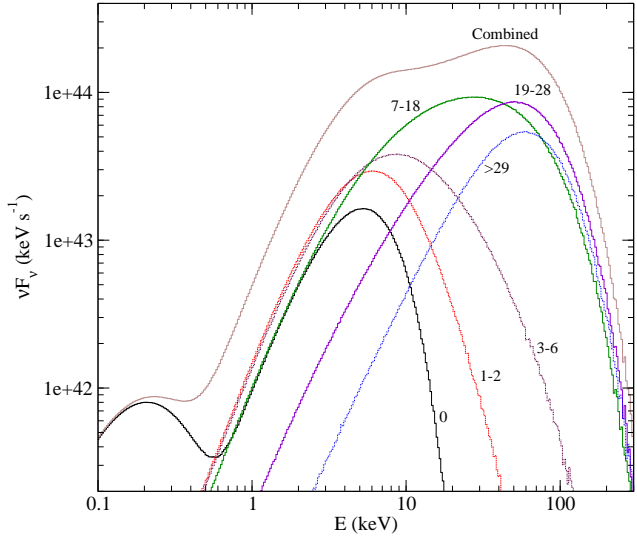}
\caption{Evidence of hardening of spectra with the number of scattering. The seed photons are emitted from the NBOL 
and Keplerian disc and are Comptonized by CENBOL. 
Here, halo accretion rate $\dot{m_h}$ is $0.5$ and disc accretion rate 
is $\dot{m_d}$ is $0.1$. The photons are binned based on the number of scatterings they underwent before emerging out of the system. The binning was 
done for $0,~1-2,~3-6,~7-18,~19-28,~29$ or higher number of scatterings. These numbers are written beside the corresponding curves 
to give an idea of how variously scattered photons contribute to the spectra. The combined spectrum is also plotted.}
\label{fig3}
\end{figure}
\end{center}

We first report variation of spectra w.r.t. the number of scatterings in CENBOL. A considerable fraction of the seed photons 
emitted are intercepted by CENBOL. This interception and subsequent Comptonization depend on the density and 
temperature along the path of the photon. These parameters are however governed by the accretion rates. 
In case of a black hole, matter is advected in through the event horizon and the efficiency of radiation is around $6\%$, 
leading to a large, though notional, upper limit ($\sim 16~\dot{M}_{EDD}$) of the Keplerian disc 
accretion rate. In the case of a neutron star, the hard surface ensures the stopping of the flow at the surface 
and the radiation decreases the upper limit of maximum acceptable accretion rate. 
For our calculations, we have strictly kept the upper limit at $1.0~\dot{M}_{EDD}$. 
The density is, thus, lower than that around a black hole, leading to a lower number of scatterings of the photons 
emitted by the Keplerian disc. The seed photons emitted by the NBOL, however, are exposed to the 
densest region of CENBOL first and that leads to a significant number of scattering. As a result, 
the contribution to hard X-Rays from seed photons that originated from the disc is smaller 
compared to those emitted from NBOL and the overall spectra, in hard states, is expected to be softer 
than the hard state spectra of a black hole. The observational differences are reported by Gilfanov (2009). 
We plot the variation of the combined spectra for the case C3 (from Table \ref{table:parameters}), 
w.r.t. number of scattering in Fig. \ref{fig3}. The binning is done for scattering number 
$0$ (no scattering), $1-2$, $3-6$, $7-18$, $19-28$, $29$ and above. The overall spectra are 
also drawn. What is clear is that the photons from the Keplerian disc are 
not scattered much. The lower number of scatterings (including those emitted without scattering)
of NBOL photons produced the first hump at around $6$ keV while higher number of scatterings which 
are effectively close to the hot region of CENBOL (i.e., centre), produce the hump at $\sim 45$ keV.

\begin{center}
\begin{figure}
\includegraphics[scale=0.35]{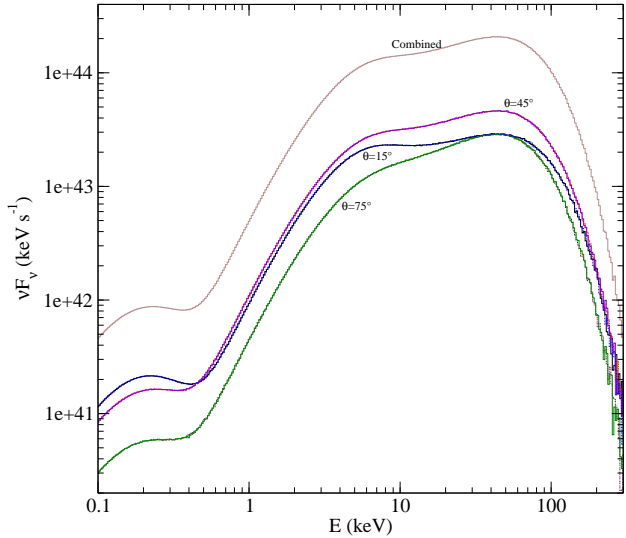}
\caption{Evidence of hardening of spectra with the observing angle. The seed photons are emitted from the NBOL 
and Keplerian disc and are Comptonized by CENBOL. 
Here, halo accretion rate $\dot{m_h}$ is $0.5$ and disc accretion rate is $\dot{m_d}$ is $0.1$. 
Photons are binned according to the direction of observation. 
All the angles are measured w.r.t. the rotation axis (z-axis). The number beside each plot shows the corresponding average angle 
for each bin (in degrees).} 
\label{fig4}
\end{figure}
\end{center}

\begin{center}
\begin{figure}
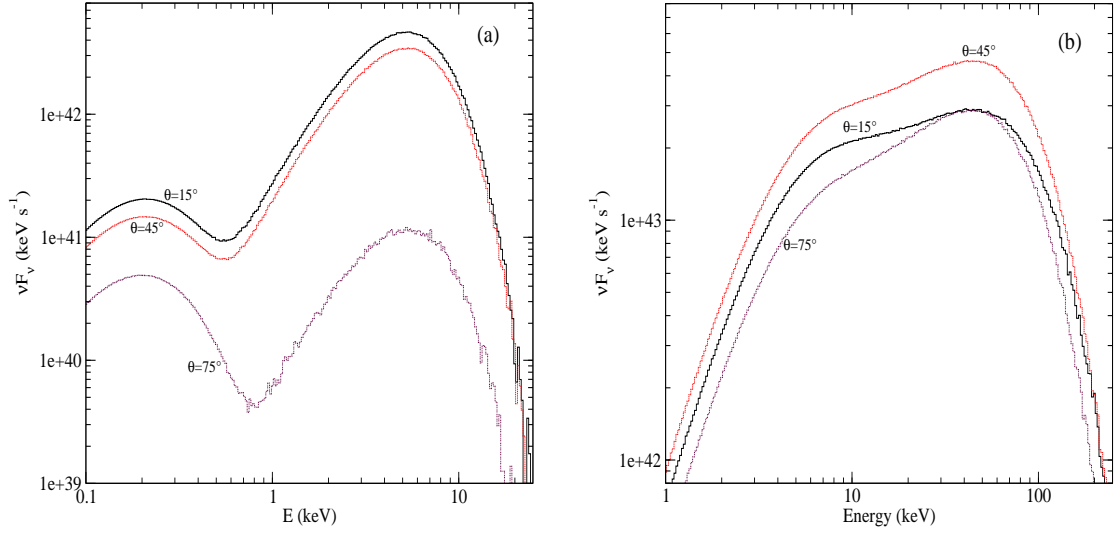

\includegraphics[height=7.0cm,width=7.0cm]{figure5a.eps}
\hspace{0.5cm}
\includegraphics[height=7.0cm,width=7.0cm]{figure5b.eps}
\caption{Evidence of hardening of spectra with the observing angle. 
The seed photons are emitted from the NBOL and Keplerian disc and are Comptonized by CENBOL. 
The flow parameters are for Case C3 of Table 1. The photons are binned according to the direction 
of observation. All the angles are measured w.r.t the rotation axis (z-axis). 
The numbers beside each plot shows the corresponding average angle bins (in degrees). 
In $(a)$, only the photons which escape without being scattered are binned. 
The angle dependency arises out of the injection direction and geometry of CENBOL. 
In $(b)$, all the photons that underwent at least one scattering are clubbed together.}
\label{fig5}
\end{figure}
\end{center}

The spectral properties of any compact object in space, be it a neutron star or a black hole, depend on the inclination 
angle between the objects line of sight and direction in which the photon is emitted. In presence of an accretion 
disc which emits photons maximally along the local normal, the received flux is maximum when the disc is seen end-on and 
minimum when seen edge-on. 
In order to check the validity of any spectral model, it has to be compared with observational data and angle dependency 
of spectra is to be studied to achieve that. We plot the variation of spectra when the photons are binned 
w.r.t. the average direction of observation, viz., $0^{\circ}-30^{\circ},~30^{\circ}-60^{\circ},~60^{\circ}-90^{\circ}$. 
In the system, both the photon sources emit photons symmetrically w.r.t. the equatorial plane or the angle $\theta=90^{\circ}$. 
The structure of CENBOL has the same symmetry as well. As no advection velocity or spin effects are included to derive
CENBOL properties, the emergent Comptonized spectra are expected to have the symmetry of the system. 
Thus in Fig. \ref{fig4}, we indicate the spectra of photons along the average bin angle in the bins of the first quadrant only.
The peak flux of unscattered spectra from the Keplerian disc decreases with increase of angle $\theta$.
In case of the radiation from NBOL, the peak flux direction is decided by the peaks of the 
double Gaussian distribution of emitted flux used in the simulation (Case C3) are along $\theta_P^-=54.94^{\circ}$ in the first quadrant.
The flux is modulated further due to the interception by CENBOL. The geometry of the post-shock region blocks 
the escape of unscattered photons emitted at high values of $\theta$. This is reflected in 
Fig. \ref{fig5}(a). The low energy peaks are due to photons emitted from the Keplerian disc and the 
high energy peaks are formed by photons from NBOL. The scattered photons, however, with the increase of 
the number of scatterings, lose their initial directional distribution and are re-distributed almost isotropically. Although, the presence of the disc and the neutron star, both of which recapture scattered photons, lowers the number of photons 
received at around angle $\theta=90^{\circ}$. The net flux is highest at $ \sim 45^\circ$ since the flux of NBOL 
is high at $\theta=\theta^-_P=54.94^{\circ}$ and still has some direct effect. The remnant effect of the double 
Gaussian enhances the flux along this direction. The Fig. \ref{fig5}(b) showcases this.  

\begin{center}
\begin{figure}
\includegraphics[scale=0.35]{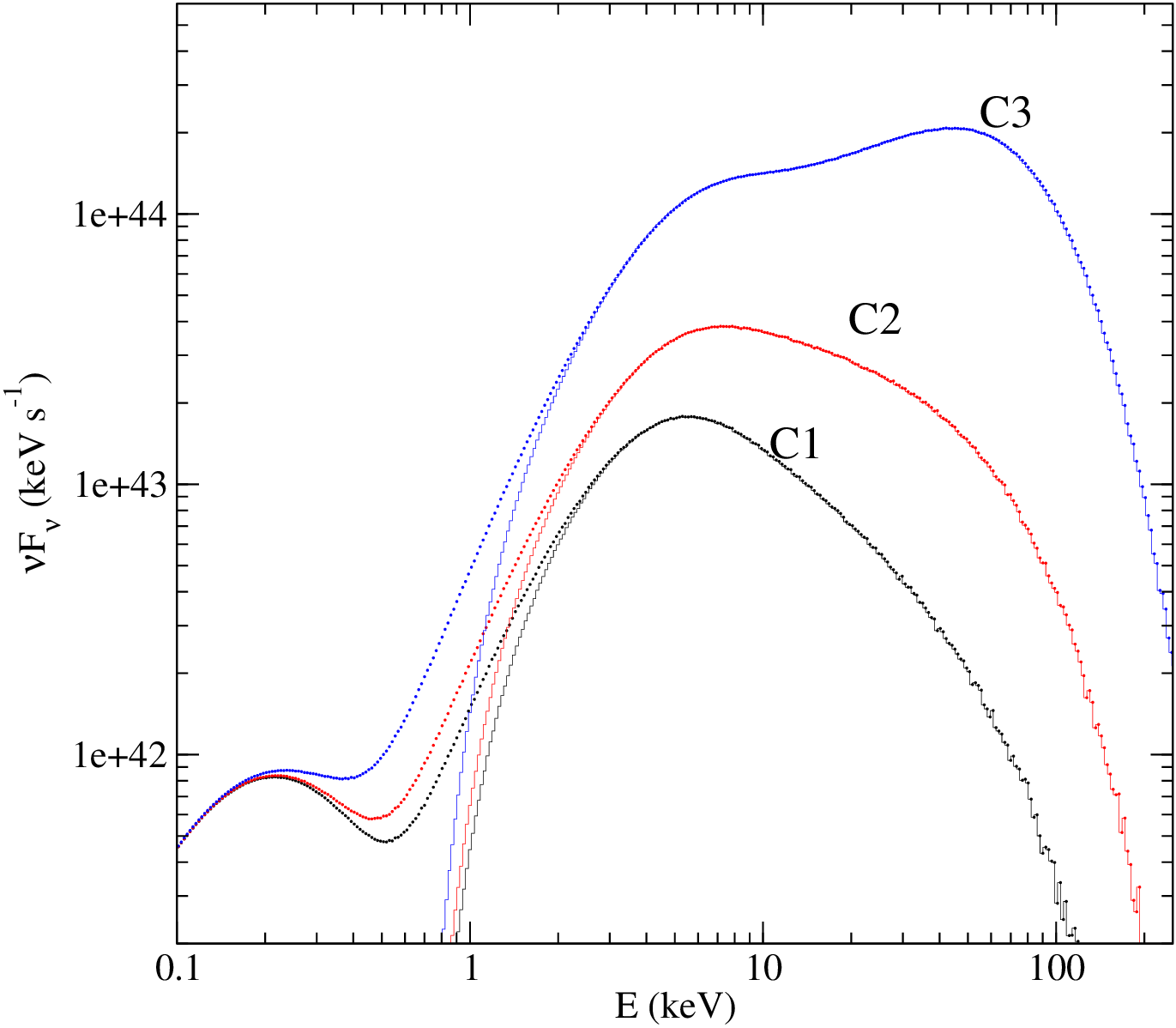}
\caption{The variation of combined spectra from NBOL, KD and the Comptonized photons 
as $\dot{m_h}$ is varied from $0.1$, $0.2$ and $0.5$. Here, $\dot{m}_d=0.1$ (Cases C1 to C3). The dotted lines show our computed spectra as are emitted from the disc while the solid lines show our spectra after absorption through ISM as they reach us. Here, we used photoelectric absorption due to hydrogen atoms, with $n_H=4 \times 10^{22}~cm^{-2}$. The cases C1, C2 and C3 are shown in black, red and blue, respectively (online version).}
\label{fig6}
\end{figure}
\end{center}

\begin{center}
\begin{figure}
\includegraphics[scale=0.35]{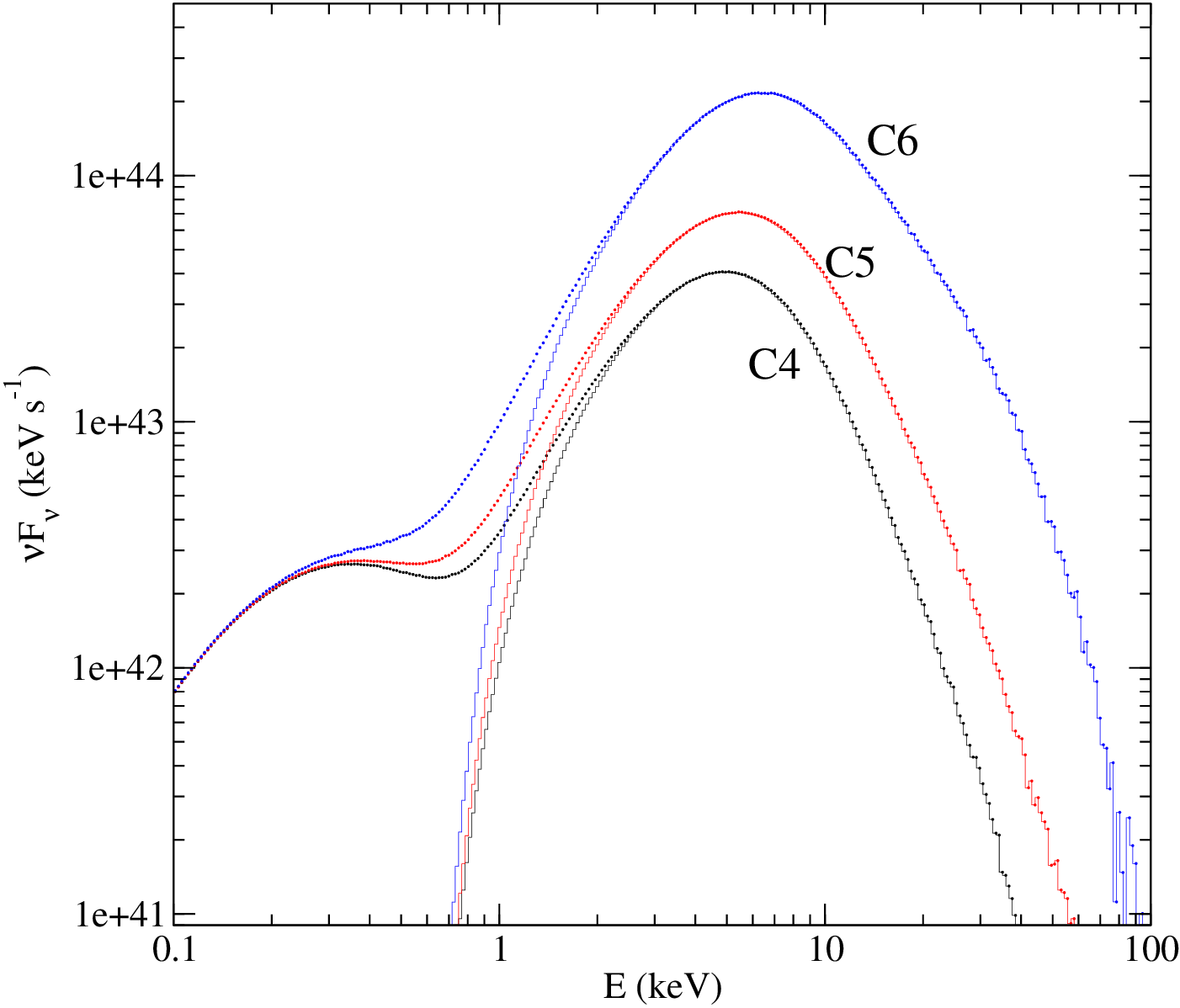}
\caption{Variation of combined spectra from NBOL, KD and the Comptonized photons as $\dot{m_h}$ is varied from $0.1$, $0.2$ and $0.5$. Here, $\dot{m}_d=0.2$. (Cases C4 to C6). Dotted lines show our computed spectra as are emitted from the disc while the solid lines show our spectra after absorption through ISM as they reach us. Here, we used photoelectric absorption due to hydrogen atoms, with $n_H=4 \times 10^{22}~cm^{-2}$. The cases C4, C5 and C6 are shown in black, red and blue, respectively (online version).}
\label{fig7}
\end{figure}
\end{center}

\begin{center}
\begin{figure}
\includegraphics[scale=0.35]{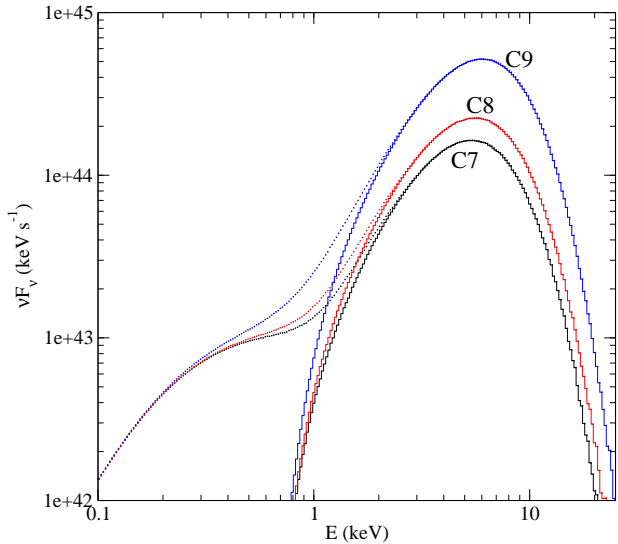}
\caption{Variation of combined spectra from NBOL, KD and the Comptonized photons as $\dot{m_h}$ is varied from $0.1$, $0.2$ and $0.5$. Here, $\dot{m}_d=0.5$. (Cases C7 to C9). Dotted lines show our computed spectra as are emitted from the disc while the solid lines show our spectra after absorption through ISM as they reach us. Here, we used photoelectric absorption due to hydrogen atoms, with $n_H=4 \times 10^{22}~cm^{-2}$. The cases C7, C8 and C9 are shown in black, red and blue, respectively (online version).}
\label{fig8}
\end{figure}
\end{center}

As shown in the context of black holes (CT95, Ghosh et al. 2009), a rise of halo accretion rate for a given disc accretion rate, 
makes the spectrum harder. Because of its hard surface, the upper limit of accretion rate for neutron stars is also
much lower than that for black holes.
This restricts the density of post-shock region leading to relatively softer spectra when compared to the spectra of black holes. 
Of course, a major factor is the abundance of seed photons from NBOL itself. 
Figures \ref{fig6}, \ref{fig7} and \ref{fig8} show how, in each case, the spectrum is affected by the increase of the 
halo accretion rate. In order to account for the cooling due to photons emitted by disc, we decreased the temperature of CENBOL self-consistently with the increase of $\dot{m}_d$
(see, Table 1). In Fig. \ref{fig8}, the spectrum is roughly a superposition of two blackbody emissions in each of the three cases. 
In Fig. \ref{fig7}, the Compton up-scattering slightly modifies the spectra but they still remain in typical soft states. 
In Fig. \ref{fig6}, where $T_{CE}=25~keV$, the spectrum changes from soft to hard with the increase of $\dot{m}_h$. We take the linear domain of the $log(\nu F_{\nu})$ vs. $log(E)$ curves plotted in logarithmic scales, if present, and try to fit the data with a power-law having spectral index $\alpha$ ($\nu F_{\nu}\sim E^{-\alpha}$). The spectral index $\alpha$, has the values 
$0.926,~0.357$ and $-0.239$ as $\dot{m}_h$ takes the values $0.1$, $0.2$ and $0.5$, respectively.
The indices are calculated from the best fit of the spectra in the energy range $10.0$ to $20.0$ keV. We have shown the theoretical spectra for cases C1 to C9, in Figs. \ref{fig6}, \ref{fig7}, and \ref{fig8}, with dotted lines and the spectra 
after absorption through ISM are plotted with solid lines. Cases C1, C4, and C7 are in black, C2, C5, and 
C8 are in red and C3, C6 and C9 are in blue (online version). Please note that unlike some models (e.g., White et al. 1986, Mitsuda et al. 1989) we do not give emphasis on the NBOL temperature or the decrease of the total luminosity. Rather, our harder  states are primarily achieved due to hotter CENBOL 
with higher advective halo rates. 

\begin{center}
\begin{figure}
\includegraphics[height=8.0cm,width=8.0cm]{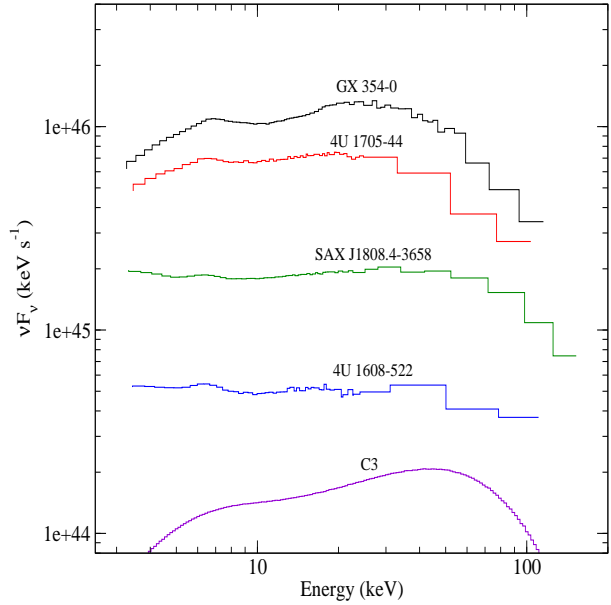}
\caption{The spectra of a few weakly magnetised neutron stars, namely, 
4U 1705-44, 4U 1608-522, SAX J1808.4-3658 and GX 354-0. For comparison, we put the spectrum of Case C3 (without the iron line emission)
of our simulation to show that we generally reproduce the features. These observed spectra were obtained by RXTE observations and 
are adapted from Gilfanov (2010).}
\label{fig9}
\end{figure}
\end{center}

The absorption due to the presence of interstellar medium modifies the low energy spectra considerably. In the hard states, when the disc accretion rate is low, the multi-colour blackbody component is hardly observed as a separate peak due to the heavy absorption below 1 keV. This was reflected in the Figs. \ref{fig6} to \ref{fig8}. After these considerations, the spectra corresponding to Case C3, looks similar to hard state 
spectra of neutron stars (see Fig. \ref{fig9}, adapted from Gilfanov 2010). The spectrum of neutron stars in hard states is relatively softer than those of black holes, because of the upper bound of maximum accretion rate. The spectra of a number of weakly magnetised accreting stars were chosen to highlight that fact in Fig. \ref{fig9}. They also have the characteristic iron line emission around 6.5 keV. Apart from that feature, the observed spectra of neutron stars, as reported in Gilfanov (2010), are similar to our Case C3, which does not include the iron line emission. It can also be concluded that the variation of the parameters of our model, e.g., $T_{NS},~T_{CE},~X_s,~R_{comp},~\dot{m}_d~and~\dot{m}_h$ can reproduce the observed spectra when suitable normalization is used. In this paper, we only report the variation of spectra with accretion rates. The variation of other parameters and comparison with observed spectra would be reported elsewhere. 

To check the variation of the effective geometry of the Compton cloud, CENBOL, as discussed in Section 4.3, we varied the accretion 
rates and the central temperature $T_{CE}$ of the CENBOL and observed how the Compton scattering changed 
the temperature profile. For $T_{CE}$ values reported in Table 1, the effect was insignificant and are 
not shown here. But, the cooling mechanism is more prominent when the central temperature is high. To 
showcase these effects we chose the case where $T_{CE}=250~keV$, $X_s=46.8~r_S$. We are reporting four 
cases here where the halo accretion rate is varied from $0.3$ to $0.9$, as shown in Fig. 10(a) through 
10(d). As the accretion rates increase, so does the temperature of the NBOL, the cooling is more efficient. 
Not only that, the effective temperature closer to NBOL decreases more than the one nearer to the disc, 
as can be seen from the shifting of the peak towards the disc in the Figure 10(a-d). From these figures, 
one can see that the effective geometry of the CENBOL is similar to the ones proposed by TSS14 for the 
flaring branches. The temperature of cloud near the disc are in between $\sim30$ keV to $\sim65$ keV for the 
cases studied here, which are in the same ballpark figure as found from COMPTB model analysis in TSS14 
and STSS15.

\begin{center}
\begin{figure}
\includegraphics[height=5.5cm,width=4.0cm]{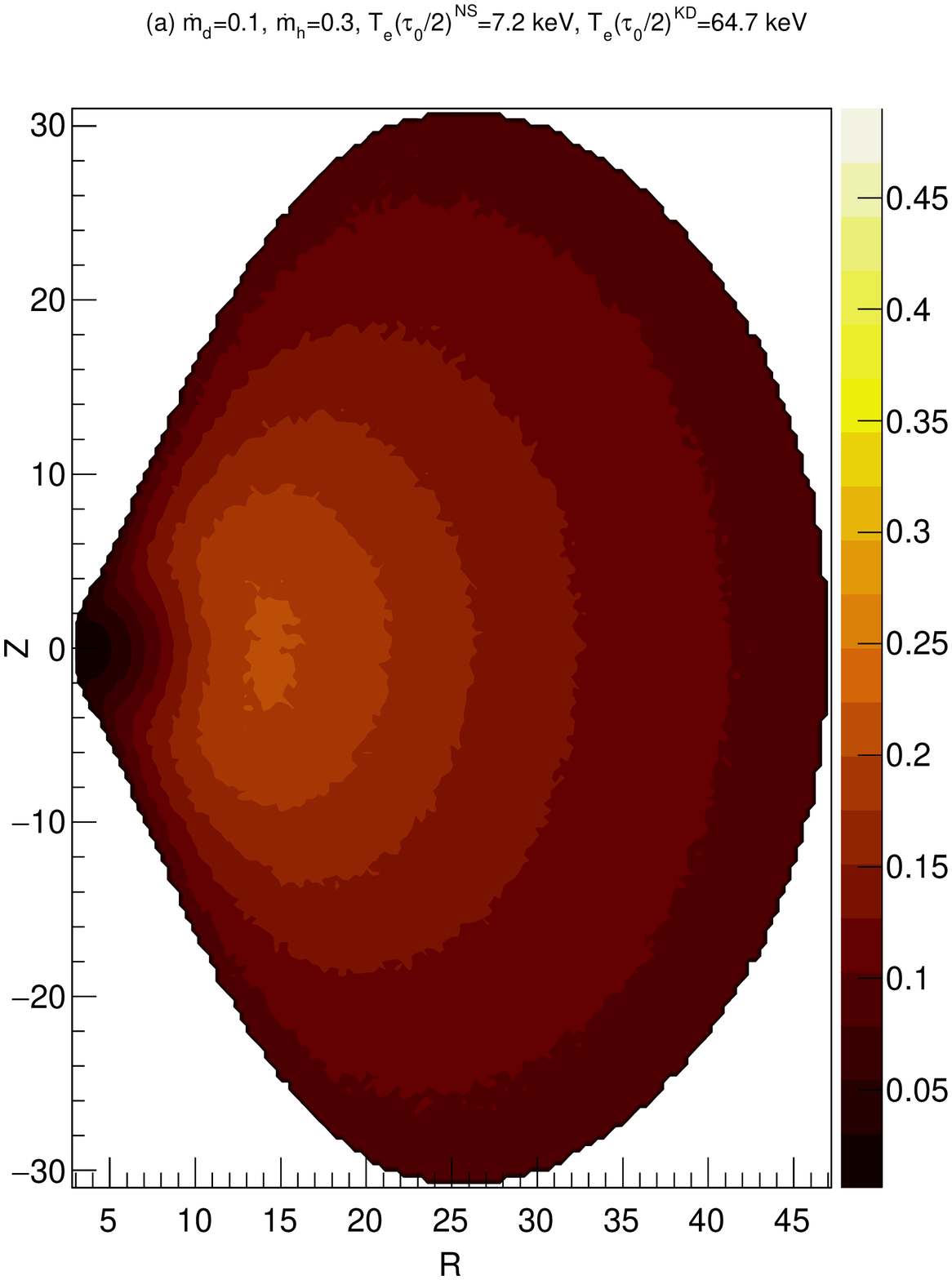}
\includegraphics[height=5.5cm,width=4.0cm]{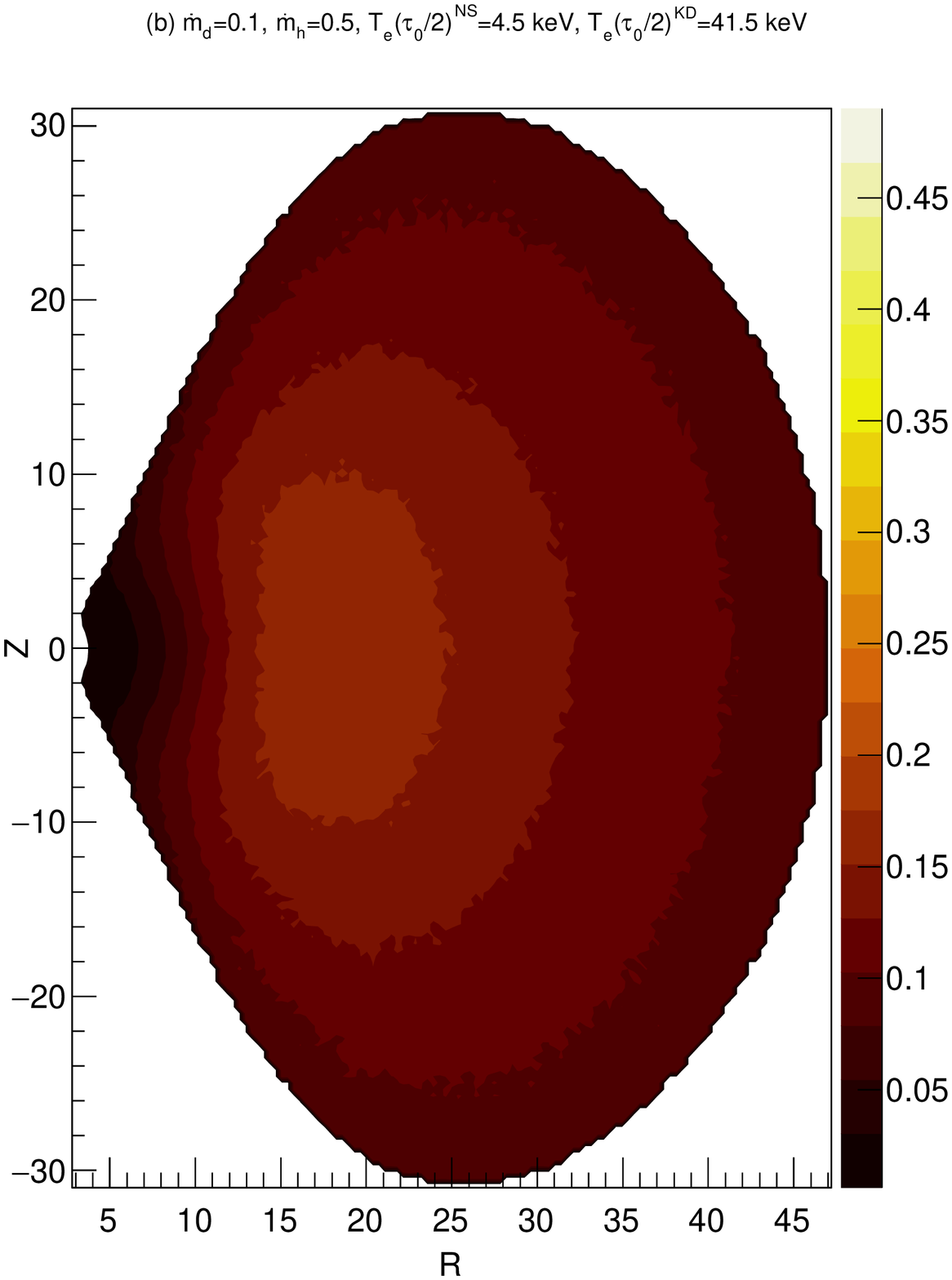}
\includegraphics[height=5.5cm,width=4.0cm]{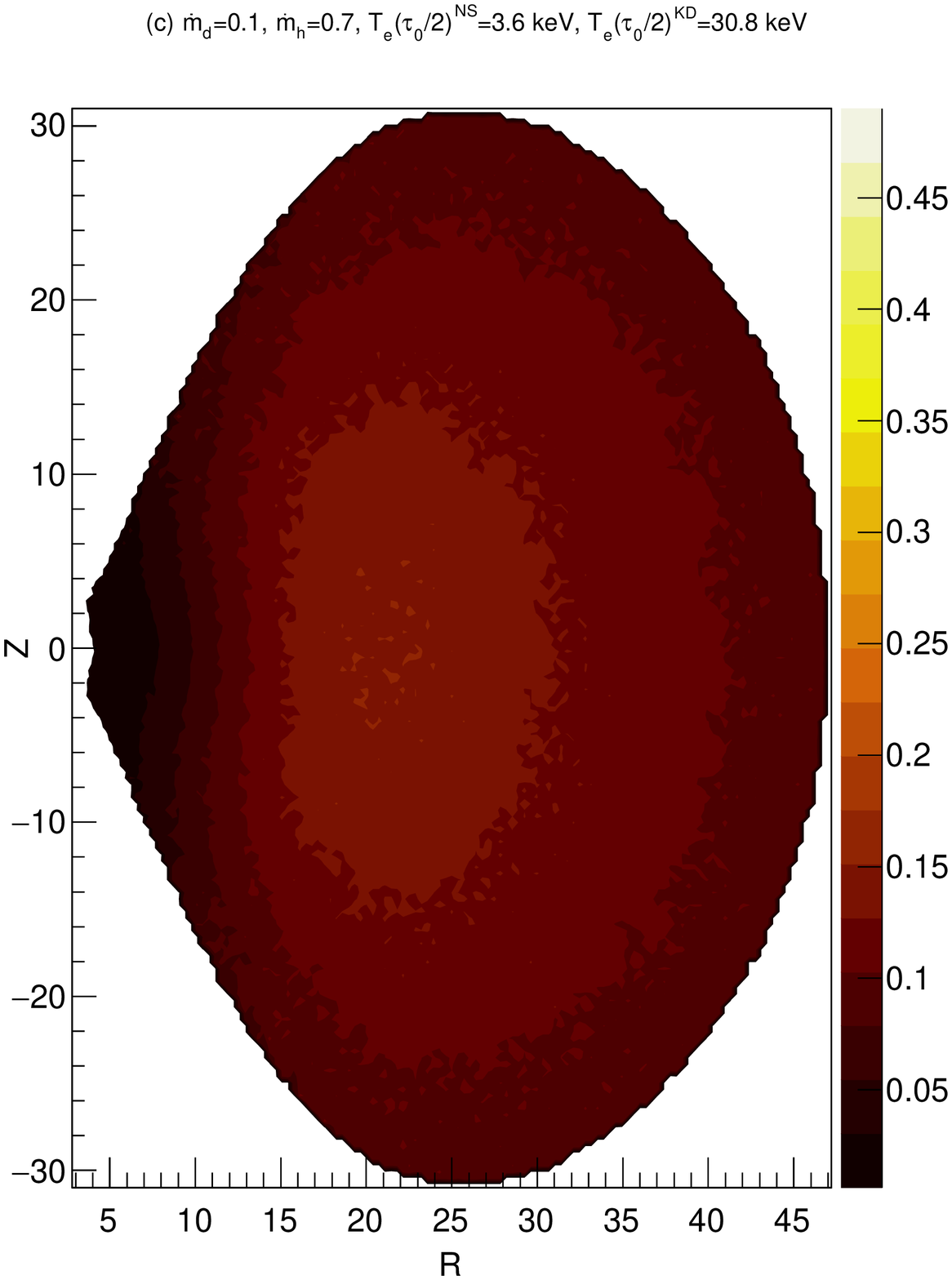}
\includegraphics[height=5.5cm,width=4.0cm]{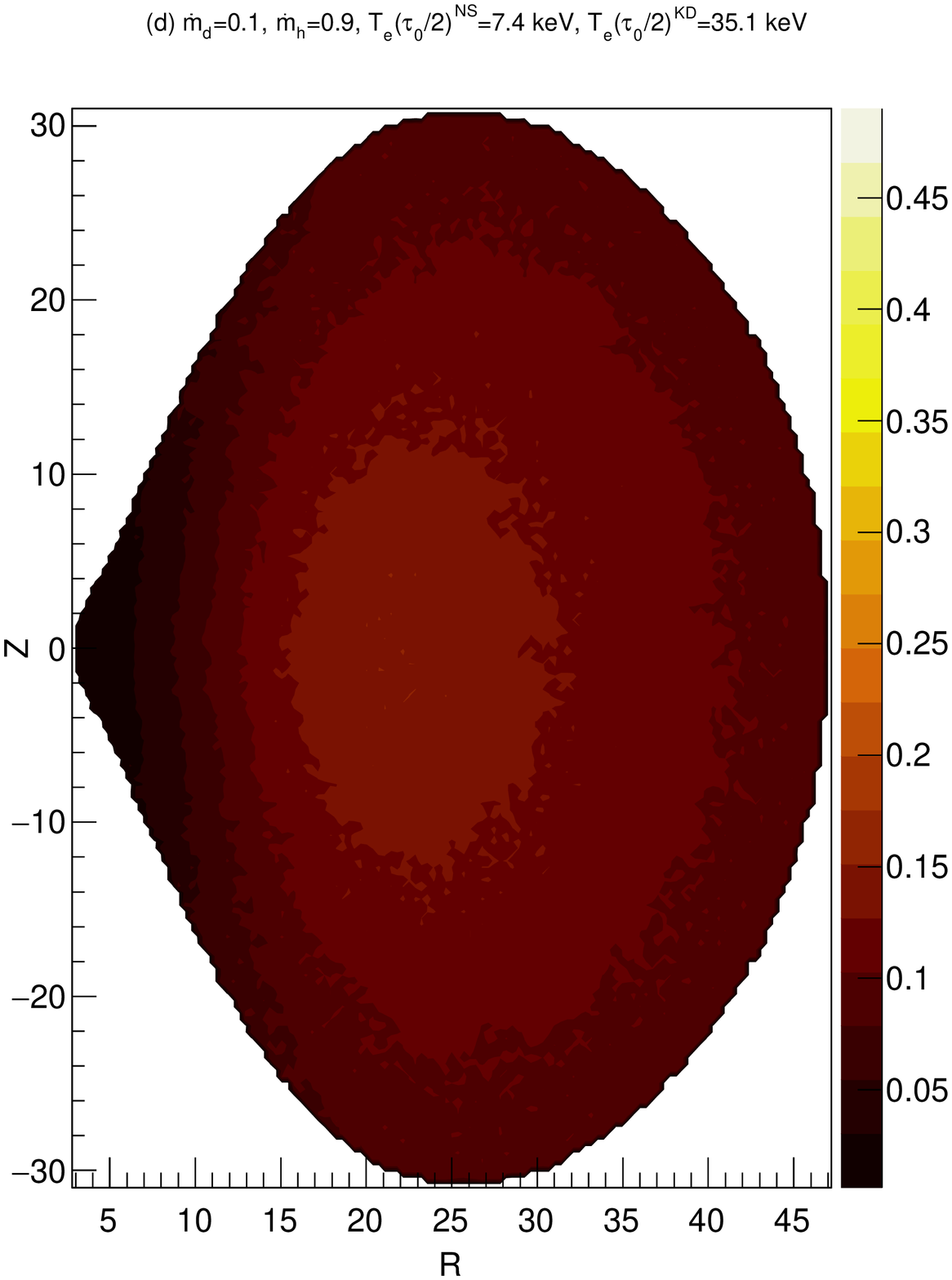}
\caption{The temperature contours after Compton cooling, from the simulation (for initial 
$T_{CE}=250.0~keV$, $X_s=46.8~r_S$). The temperatures are written in dimensionless unit $(kT_e/mc^2)$. 
From left to right, the halo accretion rate is increased from $0.3$ to $0.9$. For (a) to (c), with 
the increase of $\dot{m}_h$, the temperature of CENBOL closer to NBOL, $T_e(\tau_0/2)^{NS}$ decreased 
from $7.2$ keV to $3.6$ keV. The temperature of CENBOL closer to disc, $T_e(\tau_0/2)^{KD}$, also 
decreased from $64.7$ keV to $30.8$ keV, but remain greater than the corresponding $T_e(\tau_0/2)^{NS}$. 
In the case (d), where the angle $\theta_*$ was $\pi/2$, a large fraction of blackbody photons escaped 
the system without scattering and hence the effect of cooling, although present, is less than the 
case (c). The modified contours show the effective region of Compton scattering and are similar to 
the proposed geometry of TSS14.}
\label{fig10}
\end{figure}
\end{center}

\section{Discussions and Conclusions} 

In the literature, much studies have been done of the advective flows around black holes and 
their spectral properties. These solutions typically consist of two components, with a Keplerian disc
flanked vertically by an advective component. Moreover, the inner part of the advective component forms a 
centrifugal barrier and the post-barrier region (CENBOL) was found to behave as the Compton cloud producing typically the power-law  with
exponential cut-off. CT95 has demonstrated that the spectrum changes from soft to hard as the relative importance of 
accretion rate of the Keplerian component vis-a-vis the accretion rate of the halo is decreased. 
Since the segregation of the advective flow into a Keplerian and sub-Keplerian components could take place very far out, there is 
no {\it a priori} reason why a neutron star accretion could not also manifest into the same two types of components, especially 
that very far away the flow has not knowledge of the nature of the compact object.

In the present paper, we used the successful paradigm of TCAF in the context of a weakly magnetised neutron star accretion flow. 
The flow properties were changed on the neutron star surface due to modification of the inner boundary condition and  
a blackbody emitting normal boundary layer (NBOL) is created on the neutron star surface, in addition to the CENBOL in TCAF. 
Due to the extra source of seed photons, the CENBOL became cooler easily and the spectra appeared to be softer for the same input flow parameters. The behaviour of CENBOL is similar to the 
Compton cloud invoked phenomenologically by Titarchuk et al. (TSS14).
In our analysis, we restricted our studies within one Eddington rate. However, the general conclusion that the spectrum is hardened
with the increase in halo accretion rate remains valid. 

The resulting spectrum has several features arising out of the specific flux emission properties of the
normal boundary layer, namely, NBOL. The radiation dynamics of NBOL with Keplerian disc reveals (IS99) that
the maximum flux emitted from the NBOL may not be really along the equatorial plane, but from an angle ($\theta_*)$. 
This, together with the fact a CENBOL intercepts more photons from the NBOL than the Keplerian component, produces double 
hump patterns in the spectra. Photons from NBOL are inverse Comptonized more efficiently and thus
dictate the spectra to a greater extent.  We also studied how the spectra would change with 
the observing angle and found that with increasing inclination angle, the 
spectra is indeed hardened, a result also valid for black hole accretion (GCL09; GGCL10). However, unlike a black hole
accretion, in the present scenario, the radiation could be maximum at an intermediate viewing angle (e.g., $\sim 45$ in Fig. 4) 
and not necessarily along the polar axis.

The observed spectra from neutron star candidates, especially those which have weak magnetic fields, are found to have similar
shapes as found in the present paper thus confirming the general idea that TCAF can be used in the spectral study 
of neutron stars as well. For instance, the hard spectra of case C3 (Fig. \ref{fig6})
are similar to those of several neutron stars (Gilfanov, 2009). Similar results have also been reported by Lin et al. in 2007. Thus the motivation of our exercise to check if 
the studies of the black hole and neutron star spectra could be carried out under a common framework is well justified. 

Chakrabarti  (1997) expanded CT95 work on TCAF to establish that the advective component of accretion (the sub-Keplerian halo)
is essential to produce the hotter CENBOL. Otherwise, if the transition layer is produced solely from the
Keplerian disc and its optical depth and temperature were calculated self-consistently,
the Comptonization would be sufficient to cool it down. Otherwise, one can use the latter quantities as
free parameters without explicit use of the second component
as in Titarchuk et al. (2014) who reproduce the spectra very satisfactorily.
In the present paper, we kept in mind the inter-relationship among the flow parameters and
thus nine cases have been put in three main groups of increasing Keplerian accretion rate.
Our NBOL cools the inner CENBOL rapidly. So in a way, the CENBOL behaves like the TL of TSS14.

It has been reported from observations that the photon index of COMPTB model for 
the $\nu F_{\nu}~vs~h\nu$ spectra reaches a saturation value of $\Gamma \sim 2$, for the Comptonized 
spectra of a neutron star.
The spectra from the disc's Comptonized components, however, show no such saturation in general.
In ST11, ST12 the spectra was fitted with $Bbody+COMPTB+Gaussian$, where the Bbody was related to the
 disc emission. In STF13 two $COMPTB+Bbody$ components were used, where both the photon indices show 
stability around the value 2. In TSF13 the cloud temperature was seen to vary from 2.9 keV to 21 keV 
without any significant change in spectral slope of COMPTB spectra due to NS, but the normalization 
decreased by a factor of eight. TSS14, STSS15 have shown the saturation of spectral index of COMPTB 
(for NS only) w.r.t the variation of the temperature of the Compton cloud. The second COMPTB component 
showed a two-phase behaviour: in HB-NB, the photon index was around 2, but in FB, the photon index decreased and 
had values $1.3<\Gamma<2$. It was stated that the spectrum at the FB is determined by high radiation 
pressure from the NS surface. Burke et al. (2017) also reported the constancy of Comptonization parameters, which reflects saturation of photon index at around $\Gamma \sim 2$. In all these cases, the illumination factor f, that controls the amount of Comptonization by the Compton cloud, underwent significant changes, which shows that the geometry or the size of the Compton cloud was changing with spectral states. In STS16, it was shown that the second COMPTB component (for the 
Comptonized spectra of the disc) showed variations around
$\Gamma \sim 2$. $\Gamma$ went below $2$ when the disc temperature is reduced from 1.1 keV to 0.8 keV,
implying an expansion of Compton cloud. A simultaneous increase of the cloud temperature (of the outer part)
was also observed. These phenomenological results can be very well understood by varying
halo accretion rate which appears to be the key controlling factor here.
With the increase of $\dot{m}_h$, more hot electrons are supplied, resulting
in the expansion of CENBOL and spectral hardening (CT95, GGC14). We explore the variation of spectra
with the halo accretion rate for such cases. The results are consistent with observed results.

The proposed geometry of the Compton cloud in TSS14, for the flaring branch shows a hotter outer TL and a cooler inner TL. When we consider the effects of cooling within the Monte Carlo simulation and modify the temperature of the CENBOL, a similar profile is obtained, as shown in Fig. 10(a-d). The peak of the distribution shifts towards the disc as accretion rate is increased, for a given initial set of temperature and shock location. The temperatures obtained by us are also consistent with the observed values as reported in TSS14 and STSS15.

Recently, the TCAF solution has been used to fit spectra of several black hole candidates and at the same time to extract physical flow properties and the mass of the compact object (Molla et al. 2017, 2016, Bhattacharjee et al. 2017 and references therein). 
In near future, we plan to extract physical flow parameters onto neutron stars as well.  
Similarly, we are also extending the time-dependent studies of TCAF flow with radiative transfer in order to understand real reason 
for the high and low-frequency quasi-periodic oscillations in neutron star systems. In a future work, we will compare the spectral fits using our model and fits obtained using COMPTB model.

\section*{Acknowledgement}
We would like to the thank the referee Dr. Lev Titarchuk for bringing to our attention the Comptonizing
properties of their transition layer.


\begin{thebibliography}{10}
\bibitem[]{}Barret, D., 2001, Adv. Space Res., 28(2), 307-321
\bibitem[]{}Barret, D., Olive, J-F., ApJ, 2002, 576, 391–401
\bibitem[]{}Bhattacharjee, A., Banerjee, I., Banerjee, A., Debnath, D., \& Chakrabarti, S.K., 2017, MNRAS, 466, 1372
\bibitem[\protect\citeauthoryear{Bhattacharjee, Chakrabarti, \& Banerjee}{2016}]{2016cosp...41E.189B} Bhattacharjee A., Chakrabarti S.~K., Banerjee A., 2016, 41st Cospar Proceedings, p. 189
\bibitem[\protect\citeauthoryear{Burke, Gilfanov, \& Sunyaev}{2017}]{2017MNRAS.466..194B} Burke M.~J., Gilfanov M., Sunyaev R., 2017, MNRAS, 466, 194 
\bibitem[]{}Chakrabarti, S.K., 1985, ApJ, 288, 1-6
\bibitem[]{}Chakrabarti, S.K., Jin, L., \& Arnett, W.D., 1987, ApJ, 313, 674 (CJA87)
\bibitem[]{}Chakrabarti, S.K., 1989, MNRAS, 240, 7 (C89)
\bibitem[]{}Chakrabarti, S. K., 1990a, “Theory of Transonic Astrophysical Flows”, World Scientific (Singapore) (C90a)
\bibitem[]{}Chakrabarti, S. K., 1990b, ApJ, 362, 406 (C90b)
\bibitem[]{}Chakrabarti, S.K., \& Wiita, P.J., 1993, ApJ, 411, 602-609
\bibitem[\protect\citeauthoryear{Chakrabarti \& Molteni}{1995}]{1995MNRAS.272...80C} Chakrabarti S.~K., Molteni D., 1995, MNRAS, 272, 80 
\bibitem[]{}Chakrabarti, S.K., \& Titarchuk, L.G., 1995, ApJ, 455, 623 (CT95)
\bibitem[]{}Chakrabarti, S. K., 1996, ApJ, 464, 66d4 (C96)
\bibitem[]{}Chakrabarti, S. K. 1997, ApJ, 484, 313 (C97)
\bibitem[]{}Chakrabarti, S.K., Sahu, S.A., 1997, A\&A, 323, 382–386 (CS97)
\bibitem[]{}Chakrabarti, S.K., 2016, Proceedings of the 14th Marcel Grossman meeting at Rome, Eds. R. Ruffini, R. Jantzen, M. Bianchi (World Scientific Co., Singapore (arXiv:1604.05955)
\bibitem[]{}Chatterjee, D., Debnath, D., Chakrabarti, S.K., Mondal, S., \& Jana, A., 2016, ApJ, 827, 88
\bibitem[]{}Debnath, D., Chakrabarti, S.K., Nandi, A., \& Mandal, S., 2008, BASI, 36, 151
\bibitem[]{}Debnath, D., Chakrabarti, S.K., \& Nandi, A., 2010, A\&A, 520, 98
\bibitem[]{}Debnath, D., Chakrabarti, S.K., \& Mondal, S., 2013, ASICS, 8, 85-88
\bibitem[]{}Debnath, D., Chakrabarti, S.K., \& Nandi, A., 2013, AdSpR, 52, 2143
\bibitem[]{}Debnath, D., Mondal, S., \& Chakrabarti, S.K., 2014, MNRAS, 440, L121
\bibitem[]{}Debnath, D., Mondal, S., \& Chakrabarti, S.K., 2015, MNRAS, 447, 1984 (D2015a)
\bibitem[]{}Debnath, D., Molla, A. A., Chakrabarti, S. K., \& Mondal, S., 2015, ApJ, 803, 59
\bibitem[]{}Di Salvo \& Stella, 2002, arXiv:astro-ph/0207219v1
\bibitem[]{}Dutta, B.G., \& Chakrabarti, S.K., 2010, MNRAS, 404, 2136-2142
\bibitem[]{}Dutta, B.G., \& Chakrabarti, S.K., 2016, ApJ, 828, 101 (8pp)
\bibitem[]{}Farinelli, R., Titarchuk, L., Paizis, A., \& Frontera, F., 2008, ApJ, 680, 602-614
\bibitem[]{}Farinelli, R., Titarchuk, L., 2011, A\&A, 525, A102 (FT11)
\bibitem[]{}Garain, S.K., Ghosh, H., \& Chakrabarti, S.K., 2014, MNRAS, 437, 1329 (GGC14)
\bibitem[]{}Ghosh, H., Chakrabarti, S.K., \& Laurent, P., 2009, IJMPD 18(11), 1693–1706 (GCL09)
\bibitem[]{}Ghosh, H., Garain, S.K., Chakrabarti, S.K., \& Laurent, P., 2010, IJMPD 19(05), 607–620 (GGCL10)
\bibitem[]{}Ghosh, H., Garain, S.K., Giri, K., \& Chakrabarti, S.K., 2011, MNRAS, 416, 959-971 (GGGC11)
\bibitem[]{}Gilfanov, M.R., 2009, 'X-Ray emission from black-hole binaries' (G09)
\bibitem[]{}Gilfanov, M.R., Sunyaev, R.A., 2014, Physics-Upsekhi, 57(4), 377-388 (GS14)
\bibitem[]{}Giri, K., \& Chakrabarti, S.K., 2013, MNRAS, 430, 2836 (GC13)
\bibitem[]{}Giri, K., Garain, S.K., \& Chakrabarti, S.K., 2015, MNRAS, 448, 3221–3228 (GGC15)
\bibitem[]{}Haardt, F., \& Maraschi, L., 1993, ApJ, 413, 507
\bibitem[]{}Inogamov, N.A., Sunyaev, R.A., 1999, Astr. Lett., 25 (5), 269-293 (IS99)
\bibitem[]{}Jana, A., Debnath, D., \& Chakrabarti, S. K., et al., 2016, ApJ, 819, 107
\bibitem[\protect\citeauthoryear{Lin, Remillard, \& Homan}{2007}]{2007ApJ...667.1073L} Lin D., Remillard R.~A., Homan J., 2007, ApJ, 667, 1073 

\bibitem[]{}Mitsuda, K., Inoue, H., Koyama, K., Makishima, K.; Matsuoka, M.; Ogawara, Y.; Suzuki, K.; Tanaka, Y.; Shibazaki, N.; Hirano, T., 1984, PASJ, 36, 741
\bibitem[]{}Mitsuda, K., Inoue, H., Nakamura, N., \& Tanaka, Y., 1989, PASJ, 41(1), 97-111
\bibitem[]{}Molteni, D., Lanzafame, G., \& Chakrabarti, S.K., 1994, ApJ, 425, 161-170
\bibitem[\protect\citeauthoryear{Molteni, Sponholz, \& Chakrabarti}{1996}]{1996ApJ...457..805M} Molteni D., Sponholz H., Chakrabarti S.~K., 1996, ApJ, 457, 805 
\bibitem[]{}Molla A. A., Chakrabarti S. K., Debnath D., Mondal S., 2017, ApJ, 834, 88 (M17)

\bibitem[]{}Molla, A. A., Debnath, D., Chakrabarti, S. K., \& Mondal, S., Jana, A., 2016, MNRAS, 460, 3163 (M16)

\bibitem[]{}Mondal, S., \& Chakrabarti, S.K., 2013, MNRAS, 431, 2716 (MC13)
\bibitem[]{}Mondal, S., Debnath, D., \& Chakrabarti, S.K., 2014, ApJ, 786, 4
\bibitem[]{}Mondal, S., Chakrabarti, S.K. \& Debnath, D., 2016, Ap$\&$SS, 361, 309
\bibitem[]{}Odaka, H., Khangulyan, D., Tanaka, Y.T., Watanabe, S., Takahashi, T., \& Makishima, K., 2013, ApJ, 767:70
\bibitem[]{}Odaka, H., Khangulyan, D., Tanaka, Y.T., Watanabe, S., Takahashi, T., \& Makishima, K., 2014, ApJ, 780:38
\bibitem[]{}Paizis et al., 2006, A\&A 459, 187–197
\bibitem[]{}Popham R., Sunyaev R. A., 2001, ApJ, 547, 355
\bibitem[]{}Pozdnyakov, L.A., Sobol, I.M., \& Sunyaev, R.A., 1983, Astrophys. Space Sci. Rev. 2, 189 (PSS83)
\bibitem[\protect\citeauthoryear{Ryu, Chakrabarti, \& Molteni}{1997}]{1997ApJ...474..378R} Ryu D., Chakrabarti S.~K., Molteni D., 1997, ApJ, 474, 378 
 \bibitem[]{}Seifina, E, Titarchuk, L, 2011, ApJ, 738, 128; (ST11)
\bibitem[]{}Seifina, E., Titarchuk, L., 2012, ApJ, 747, 99 (ST12)
\bibitem[]{}Seifina, E., Titarchuk, L., Frontera, F., 2013, ApJ, 766, 63; (STF13)
\bibitem[]{}Seifina, E., Titarchuk, L., Shrader, C., Shaposhnikov, N., 2015, ApJ, 808, 142 (STSS15)
\bibitem[]{}Seifina, E., Titarchuk, L., Shaposhnikov, N.,2016. ApJ, 821, 23; (STS16)
\bibitem[]{}Seon, K.-I., Choi, C.-S., Nam, U.-W., \& Min, K.-W., 1994, JKAnS, 27(1), 45-53
\bibitem[]{}Shakura, N.I., \& Sunyaev, R.A., 1973, A\& A, 24, 337 (SS73)
\bibitem[]{}Smith, D.M., Heindl, W.A., Markwardt, C.B., \& Swank, J.H., 2001, ApJ, 554, L41-L44
\bibitem[]{}Smith, D.M., Heindl, W.A., \& Swank, J.H., 2002, ApJ, 569, 362-380
\bibitem[]{}Sunyaev, R.A., \& Truemper, J., 1979, Nature, 279, 506-508 (ST79)
\bibitem[]{}Sunyaev, R.A., \& Titarchuk, L.G., 1980, ApJ, 86, 121 (ST80)
\bibitem[]{}Sunyaev, R.A., \& Titarchuk, L.G., 1985, A\& A, 143, 374 (ST85)

\bibitem[\protect\citeauthoryear{Titarchuk, Lapidus, \& Muslimov}{1998}]{1998ApJ...499..315T} Titarchuk L., Lapidus I., Muslimov A., 1998, ApJ, 499, 315 (TLM98)
\bibitem[]{}Titarchuk, L., Seifina, E., Frontera, F., 2013, ApJ, 767, 160; (TSF13)
\bibitem[]{}Titarchuk, L., Seifina, E., \& Shrader, C., 2014, ApJ, 789, 98 (TSS14) 
\bibitem[]{}White, N.E., Peacock, A., Hasinger, G., Mason, K.O., Manzo, G., Taylor, B.G., \& BranduardiRaymont, G., 1986, MNRAS, 218, 129-138
\bibitem[]{} Zdziarski, A.A., Lubinski, P., \& Gilfanov, M., et al., 2003, MNRAS, 342, 355
\end{thebibliography}
\end{document}